\input amssym.def
\input amssym.tex
\documentstyle[eqsecnum,preprint,aps,prd]{revtex}
\preprint{DAMTP-R95/60} 
\date{\today}

\tighten
\begin{document} 
\draft

\title{Pair Creation of Black Holes by Domain Walls}

\author{R.R. Caldwell{$^1$}, A. Chamblin{$^{1,2}$}, and G.W. Gibbons{$^1$}}
 
\address{\qquad \\ {$^1$}University of Cambridge, D.A.M.T.P.\\
Silver Street, Cambridge CB3 9EW, U.K.\\
\qquad\\
{$^2$}Institute for Theoretical Physics\\
University of California\\
Santa Barbara, California 93106-4030, U.S.A.}

\maketitle

\begin{abstract}

In this paper we study  the production of pairs of neutral and charged
black holes by domain walls, finding classical solutions and
calculating their classical actions. We find that neutral black holes
whose creation is mediated by Euclidean instantons must be produced
mutually at rest with respect to one another, but for charged black
holes a new type of instanton is possible in which after formation the
two black holes accelerate away from one another.  These new types of
instantons are  not possible in Einstein-Maxwell theory with a
cosmological constant.  We also find that the creation of non-orientable
black hole solutions can be mediated by Euclidean instantons and that
in addition if one is prepared to consider entirely Lorentzian
no-boundary type contributions to the path integral then mutually
accelerating pairs may be created even in the neutral case.  Finally we
consider the production of Kaluza-Klein monopoles both by a standard
cosmological term and in the presence of a domain wall. We find that
compactification is accompanied by the production of pairs of
Kaluza-Klein monopoles.

\end{abstract}
\vskip 0.3in

%%%%%%%%%%%%%%%%%%%%%%%%%%%%%%%%%%%%%%%%%%%%%%%%%%%%%%%%%%%%%%%%%%%
%%%%%%%%%%%%%%%%%%%%%%%%%%%%%%%%%%%%%%%%%%%%%%%%%%%%%%%%%%%%%%%%%%%
%%%%%%%%%%%%%%%%%%%%%%%%%%%%%%%%%%%%%%%%%%%%%%%%%%%%%%%%%%%%%%%%%%%
\section{Introduction} 
\label{intro}

Motivated by the recent interest in black hole pair creation in
Euclidean quantum gravity \cite{EUCLIDEAN}, we consider additional such
tunnelling processes in the presence of domain walls.  The reason we
expect tunnelling processes occur in a domain wall space-time is
because the gravitational field of a domain wall is repulsive.  An
observer near a domain wall will experience a repulsive force, and will
accelerate away from the wall. This is just the sort of
`anti-gravitational' background in which pair creation processes occur,
as with de Sitter space-time \cite{DESITTER}. In this paper we show
that tunnelling geometries do indeed exist, and compute the
probabilities for the pair creation of uncharged and charged black
holes in a domain wall space-time.  In section \ref{domwalls}, we
discuss the novel gravitational properties of domain walls and give a
cut-and-paste procedure for the construction of a domain wall
space-time.  In section \ref{domains}, we consider the tunnelling
process by which vacuum domains are created from nothing. We construct
the necessary instanton, and compute the tunnelling amplitude.  This
process is to be considered as the background process to the pair
creation of black holes. In section \ref{uncharged}, we show that
uncharged, static black holes may be pair created in the presence of a
domain wall. This result is unusual, as few other processes for the
pair creation of uncharged black holes are known. However, no instanton
exists for the creation of accelerating, uncharged, static black
holes.  In section \ref{charged}, we consider the pair creation of
magnetically charged black holes. Here we find both static and
accelerating solutions, with two unusual results. First, for a given
domain wall surface energy there exists an infinite sequence of
discrete values of the charge, describing the production of
accelerating, charged black holes. Second, under the action of certain
discrete involutive isometries, half of these configurations describe
the creation of non-orientable black holes. In section \ref{uniqueness}
we discuss the uniqueness and isoperimetry of instantons. In section
\ref{lorentzian} we discuss Lorentzian tunnelling geometries for the
processes described in this paper.  Finally, in section
\ref{conclusion}, we discuss the implications of the creation of
non-orientable black holes, the stability of domain walls against
puncture.

Throughout this paper we use units in which $\hbar = c = G = 1$.

%%%%%%%%%%%%%%%%%%%%%%%%%%%%%%%%%%%%%%%%%%%%%%%%%%%%%%%%%%%%%%%%%%%
%%%%%%%%%%%%%%%%%%%%%%%%%%%%%%%%%%%%%%%%%%%%%%%%%%%%%%%%%%%%%%%%%%%
%%%%%%%%%%%%%%%%%%%%%%%%%%%%%%%%%%%%%%%%%%%%%%%%%%%%%%%%%%%%%%%%%%%

\section{Domain Walls}
\label{domwalls}

In this section we review the properties of domain walls.  We discuss
the novel gravitational properties of a domain wall space-time,
features relevant for the calculations carried out in this paper. A
procedure for the cut-and-paste construction of a domain wall
space-time will be given.

A domain wall is a two-dimensional topological defect which forms at
the boundary between two regions of space in which a field, such as a
Higgs, has undergone the breaking of a discrete symmetry.  If $V_0$ is
the submanifold of the field configuration space on which the field
acquires a vacuum expectation value, then a necessary condition for the
appearance of a domain wall is $\pi_0(V_0) \neq 0$ which tells us that
the vacuum manifold is not connected.  The physics of domain walls is
extensively reviewed in \cite{VILENKINSHELLARD}.

We now summarize the gravitational features of
a domain wall \cite{VILENKINSHELLARD,VILENKIN,IPSER}.
An idealized, thin domain wall located at $x=0$ 
has the stress-energy tensor
\begin{equation}
T_{\mu \nu} = \sigma \delta(x) {\rm diag} (1,0,1,1) 
\label{stressenergy}
\end{equation} 
where the surface mass-density of the wall is
given by $\sigma$.  For this distributional source, the
metric is $C^0$ but {\it not} $C^{\alpha}$, for any $\alpha
\geq 1$.  While it is not possible to find a {\it static}
solution of the Einstein equations with this source term,
a time-dependent solution exists, as  
shown by Vilenkin \cite{VILENKIN} and Ipser \& Sikivie
\cite{IPSER}.  Their metric takes the form
\begin{equation} 
ds^{2} = \Big(1 - {\cal K} |x|\Big)^2
dt^{2} - dx^{2} - \Big(1 - {\cal K} |x|\Big)^2 
e^{2{\cal K} t}   (dy^{2} + dz^{2}),
\label{dwmetric}
\end{equation} 
where ${\cal K} = 2{\pi}{\sigma}$.  
As was pointed out in \cite{VILENKIN},
the gravitational field of the vacuum domain wall described
by (\ref{dwmetric}) is repulsive, because
the source (\ref{stressenergy}) violates the strong energy 
condition; an inertial observer at $x = 0$ will see test bodies
accelerated away from the wall with acceleration 
$a = {\cal K}$.  To understand this, note that the $t - x$
part of the metric (\ref{dwmetric}) is the two-dimensional
Rindler metric \cite{IPSER}. By analogy with the cases
of electric fields and a positive cosmological constant,
we would expect this repulsive force to provide a mechanism 
for a tunnelling process such as black hole pair production, 
as we find later in this paper. 

Let us examine in more detail the global structure of the 
space-time described by the line-element (\ref{dwmetric}). 
The hypersurfaces $x$= constant are {\it isometric} to a 
portion of $2 + 1$-dimensional de Sitter space-time:  
\begin{equation} 
ds^{2} = dt^{2} - e^{2{\cal K} t}(dy^{2} + dz^{2}). 
\end{equation} 
Since $2 + 1$ de Sitter space-time has the topology
${\rm S}^{2} \times {\Bbb R}$ we expect the
domain wall to have the topology 
${\rm S}^{2} \times {\Bbb R}$, which means that at each 
instant of time, the domain wall is a sphere.   
To see this more clearly, recall that in \cite{IPSER} a 
transformation to the coordinates $(T, X, Y, Z)$  
was found covering one side of the domain wall space-time,
so that the metric becomes 
\begin{equation} 
ds^{2} = dT^{2} - dX^{2} - dY^{2} - dZ^{2}.  
\end{equation}
Thus the domain wall, which in the old coordinates is a plane located at
$x = 0$, is in the new coordinates the hyperboloid
\begin{equation} 
X^{2} + Y^{2} + Z^{2} = {1\over {\cal K}^2} + T^{2}.  
\label{dwhyperboloid} 
\end{equation} 
The metric induced on this hyperboloid by the ambient Minkowski
metric is just the de Sitter metric. Therefore, the domain wall
{\sl is} a copy of $2 + 1$ de Sitter. This provides a useful way 
of thinking of this domain wall. Consider the following
topological construction: take two
copies of Minkowski space, and in each copy consider the
interior of the hyperboloid determined by equation
(\ref{dwhyperboloid}), match these solid hyperboloids to
each other across their respective boundaries; there will
be a ridge of curvature (much like the edge of of a lens)
along the matching surface, where the domain wall is
located. This is illustrated in figure \ref{figure1}.  
Thus, an inertial observer on one side of the wall will see the
domain wall as a sphere which accelerates 
towards the observer for $T<0$, 
stops at $T=0$ at a radius ${\cal K}^{-1}$,
then accelerates away for $T>0$.
With this brief introduction to the properties of
the domain wall space-time, we turn our attention to the problem of 
using this background to nucleate black holes.

%%%%%%%%%%%%%%%%%%%%%%%%%%%%%%%%%%%%%%%%%%%%%%%%%%%%%%%%%%%%%%%%%%%
%%%%%%%%%%%%%%%%%%%%%%%%%%%%%%%%%%%%%%%%%%%%%%%%%%%%%%%%%%%%%%%%%%%
%%%%%%%%%%%%%%%%%%%%%%%%%%%%%%%%%%%%%%%%%%%%%%%%%%%%%%%%%%%%%%%%%%%
\section{Creation of Domains}
\label{domains}

In order to study the pair creation of black holes in a domain wall
space-time, we first consider the creation from nothing of a closed
universe consisting of two vacua separated by a domain wall. This
process is the background to the black hole pair production.

The instanton for domain creation is the analogue of the $S^4$
instanton which mediates the creation from nothing of a de Sitter
space-time. Since the Lorentzian section of the instanton is almost
everywhere flat, and consists of the interiors of two hyperboloids in
Minkowski space-time glued back-to-back, the obvious choice for the
Riemannian section is to take two flat $4-$balls and glue them
back-to-back.  This gives a manifold topologically equivalent to
$S^4$.  Thus we obtain a `lens', owing to the ridge of curvature
running along the hemisphere at the location of the domain wall,
which is topologically like $S^4$.  This construction is illustrated
in figure \ref{figure1}.
This Riemannian section may be matched to the Lorentzian space-time
across a nucleation surface ${\Sigma}$, to describe the creation of a
closed universe.

The Euclidean action for the domain creation is
\begin{equation}
S_E = \int_M \sqrt{g} d^4 x \,
\Big[ -{R \over 16 \pi} + {1 \over 2} (\partial \phi)^2
+ V(\phi) \Big].
\end{equation}
Substituting the on shell condition
\begin{equation}
{R \over 8 \pi} = (\partial \phi)^2 + 4 V
\end{equation}
we obtain
\begin{equation}
S_E = -\int \sqrt{g} d^4 x \, V(\phi).
\end{equation}
There are no boundary terms
because the instantons we consider
are compact, having no boundary.
Also, provided the potential energy function
$V(\phi)$ is positive, the Euclidean action $S_E$
will always be negative.
For a domain wall in flat space, as a consequence of the equation of motion,
\begin{equation}
{1 \over 2} \Big({d \phi \over d z}\Big)^2 - V(\phi) = 0,
\end{equation}
where $z$ is the proper distance in the direction perpendicular to the domain
wall. The energy-per-unit-area $\sigma$ is given by
\begin{eqnarray}
\sigma &=& \int_{-\infty}^{\infty}
dz \, \Big[{1 \over 2} \Big({d \phi \over d z}\Big)^2 + V(\phi) \Big] \cr\cr
&=& 2 \int_{-\infty}^{\infty} dz V(\phi[z])
\end{eqnarray}
Thus, for domain walls in the thin wall approximation,
\begin{equation}
S_{E} = - {1 \over 2} \sigma \int \sqrt{h} d^3 x
\label{actionderived}
\end{equation}
where the integration of $\sqrt{h} d^3 x$ gives the volume of the $S^3$ ridge
on the instanton. The ridge is located at radius $r=1/(2 \pi \sigma)$ from the
center of either four-ball. Hence, the volume is
\begin{equation} 
{\rm vol}({\rm S}^3) =
2 \pi^2 r^3 = {1 \over 4 \pi \sigma^3}   
\end{equation} 
Therefore, the action for nucleating
a domain wall is  
$S_{E} = -1/(8 \pi \sigma^2)$. 
The amplitude for this process, ignoring the pre-factor is,
remembering that only half the Riemannian instanton is included in the
complex path, $\exp(-S_{E}/2)$. This amplitude increases for increasing 
domain wall size, or decreasing $\sigma$. Such behaviour is analogous
to the case of a universe containing a positive cosmological constant 
created from nothing. There, the tunnelling process amplitude grows
for an increasingly large universe, or for decreasing cosmological
constant.

%%%%%%%%%%%%%%%%%%%%%%%%%%%%%%%%%%%%%%%%%%%%%%%%%%%%%%%%%%%%%%%%%%%
%%%%%%%%%%%%%%%%%%%%%%%%%%%%%%%%%%%%%%%%%%%%%%%%%%%%%%%%%%%%%%%%%%%
%%%%%%%%%%%%%%%%%%%%%%%%%%%%%%%%%%%%%%%%%%%%%%%%%%%%%%%%%%%%%%%%%%%
\section{Uniqueness and Isoperimetry}
\label{uniqueness}

In deciding which instantons are the most important to a
given physical process one usually uses two criteria: 
(1) uniqueness, and 
(2) least action.
In fact the latter criterion is more frequently used.
Consider for example the decay of the false vacuum in
4-dimensional Euclidean space via the formation of a bubble
of true vacuum.  According to Coleman et al \cite{COLEMAN},
the $O(4)$ invariant bubble has the least action among
solutions.  In the thin wall approximation this reduces to
the isoperimetric inequality:  the 3-sphere has the least
area of any 3-surface enclosing the given 4-volume.

In fact Coleman's assumption that we should restrict attention to the
least action solution is redundant in this case.  A celebrated
theorem of Gidas, Ni and Nirenberg \cite{GIDAS} implies that the {\it
only} solutions of ${\nabla}^2{\phi} = V^{\prime}({\phi})$ on ${\Bbb
R}^4$ tending to the false vacuum at infinity are $O(4)$ invariant.
Thus independently of action considerations if we are to use a saddle
point of the classical action at all we can only use an $O(4)$
invariant one.

In the thin wall approximation the result of Gidas, Ni and Nirenberg
\cite{GIDAS} reduces to an equally celebrated result of Aleksandrov
\cite{ALEKSANDROV} on the uniqueness of embedded closed hypersurfaces
of constant mean curvature $K = g^{ij}K_{ij}$ in flat space:  They must
all be spherical.  Another closely related result is that of Ros [1]:
A sphere is the only compact hypersurface with constant Ricci scalar
embedded in Euclidean space.  The thin shell approximation is $K_{ij} =
cg_{ij}$.  The Gauss-Codazzi equation, ${\nabla}_{i}K^{ij} -
{\nabla}^{j}{K} = 0$, then imples that $c$ is a constant.  In a flat
Riemannian 4-manifold  the Ricci scalar of a hypersurface  is given by
\begin{equation} {}^{(3)}{R}  = -{K}_{ij}{K}^{ij} + {K}^{2}
\end{equation} whence ${}^{(3)}R$ is a constant.

Note that these global results may fail if we loosen our assumptions
that we have embedded hypersurfaces to allow immersions.  In our case
the assumption that we have an embedding is essential because our
domain wall must separate space-time into two regions:
self-intersections are not allowed physically.

It follows directly from the results of Aleksandrov \cite{ALEKSANDROV}
or of Ros \cite{ROS} quoted above that our construction of the
instanton mediating the birth of two domains is unique.  In other
words, our almost everywhere flat ${\rm S}^4$ is the only such almost
everywhere flat ${\rm S}^4$.  It is interesting that the corresponding
uniqueness result for the usual round Einstein metric on ${\rm S}^4$
remains elusive.  Furthermore, although we have not investigated in
detail the uniqueness of our $O(3)$ invariant thin shells in the
Euclidean Schwarzschild geometry we conjecture that they are also
unique.

%%%%%%%%%%%%%%%%%%%%%%%%%%%%%%%%%%%%%%%%%%%%%%%%%%%%%%%%%%%%%%%%%%%
%%%%%%%%%%%%%%%%%%%%%%%%%%%%%%%%%%%%%%%%%%%%%%%%%%%%%%%%%%%%%%%%%%%
%%%%%%%%%%%%%%%%%%%%%%%%%%%%%%%%%%%%%%%%%%%%%%%%%%%%%%%%%%%%%%%%%%%
\section{Creation of Uncharged Black Holes}
\label{uncharged}

We now analyse the problem of creating uncharged black holes in
the presence of a domain wall. First, we construct the instanton
for the nucleation of a pair of static black holes. Second,
we demonstrate that no such instanton exists for accelerating
black holes in this background space-time. Third, we compute the
probability for the nucleation of a pair of static black holes.

Let us address whether it is possible to create a pair of black holes
which are in static equilibrium relative to the domain wall.
Na{\"\i}vely, we would expect to be able to fine tune the black hole
mass $m$ and domain wall surface energy ${\sigma}$ so that the
repulsive force of the wall exactly balances the attractive force on
the black holes.  Indeed, this is always possible and there exists a
Euclidean instanton which we can use to estimate the probability that
such a pair of `finely tuned' black holes would be created in the
presence of the wall.

To obtain the instanton for static black hole nucleation, we first
construct the Lorentzian section. In order to obtain a non-zero 
probability, we require a spatially closed universe whose
3-volume is finite.  This guarantees that the total
energy at the instant of nucleation vanishes. 
To obtain such a space-time, take two copies of the
full vacuum Kruskal manifold, each of which has two asymptotically flat
regions. Cut each along two static timelike hypersurfaces of the same
radius, one in each asymptotically flat region outside each hole,
discarding the exteriors. This procedure is illustrated
in figure \ref{figure2}.
We obtain a space-time with closed
spatial sections having topology 
${\Bbb R} \times {\rm S}^1 \times {\rm S}^2$ by
identifying across the two static timelike hypersurfaces.  The
result is a space-time containing two domain walls and two domains
containing a black hole in each. Moreover this space-time, with
its identifications, has a Riemannian section.

To obtain this Riemannian section, we start with the usual Riemannian
section of Schwarzschild with mass $m$.  There the topology is
${\rm S}^2 \times {\Bbb R}^2$, where 
the ${\Bbb R}^2$ factor looks like a cigar.  We snip each
cigar along the radius $r = 3 m$, where 
$m = (6 \sqrt{3} \pi \sigma)^{-1}$, 
corresponding to the location of the domain wall. Next, we  
graft the two manifolds together along the surface where we made the
cut.  The resulting surface, which is rather like a `baguette',
having topology ${\rm S}^{2} \times {\rm S}^{2}$ with a `ridge' at
the domain wall, is illustrated in figure \ref{figure2}. 
One may check that for
a radius $r = 3 m$ and no other,
\begin{enumerate}
\item 
the hypersurface $r$ is totally umbilic, that is the second
fundamental form $K_{ij}$ is proportional to the
induced metric $g_{ij}$ on the domain wall world sheet
\item 
the  discontinuity in the second fundamental form on the hypersurface
$r=3m$  is $[K_{ij}]_\pm = 4 \pi \sigma g_{ij}$.
\end{enumerate}
Thus, if $r=3m$, the metric satisfies the Israel
junction conditions and is therefore a {\it bona fide}
domain wall space-time.

The Riemannian section now has topology $S^2 \times S^2$, containing a
single domain wall and two `bolts', that is two $2-$spheres on which
the Killing field ${\partial \over  \partial \tau}$ vanishes.  The
Riemannian and Lorentzian sections are joined together across the
nucleation surface $\Sigma$ which has topology $S^1 \times S^2$.  The
nucleation circle is located on the `baguette' along $\tau = 0$ 
{\sl and} $\tau = 4 \pi m$.  We now turn to the equations of motion 
of the instanton.

The above construction for the nucleation of a pair of static black
holes in the presence of a domain wall is a special case of the work
of Hiscock \cite{HISCOCK} and Berezin et al \cite{BEREZIN} 
on $O(3)-$invariant thin wall bubble
nucleation. Applying the Israel junction conditions to the
domain wall interface joining the two Euclidean black hole space-times, 
we obtain the equation of motion \cite{HISCOCK} 
\begin{equation} 
\sqrt{f - {\dot r^{2}}} = 2\pi \sigma r,
\label{eqnofmotion}
\end{equation} 
for a wall located at radius $r$.
Here $f = 1 - 2m/r$, the $g_{00}$ Euclidean
Schwarzschild metric coefficient in $(\tau,r)$ coordinates, and
$\cdot \equiv f^{-{1\over 2}} {\partial \over \partial \tau}$.  This
equation may have the interpretation as describing a fictitious particle
moving under the influence of a potential 
\begin{equation} 
V(r) = f^2 - (2 \pi \sigma r)^2.  
\end{equation} 
In the present case, the equation describes the motion of the domain
wall relative to the Euclidean black holes.
The general solution of (\ref{eqnofmotion}) is periodic,
but there is a solution
for which  $r=$constant and the energy is
zero, corresponding to $\partial V/\partial r = 0$.  This
has an infinite period, and occurs at
$r = 3 m$; it is the static domain wall.  

We may also look for accelerating solutions  
describing the creation of accelerating black holes.
The solution of the Euclidean equations of motion is
periodic in $\tau$.  The period $\beta_w$ is obtained by evaluating
the line integral $d\tau$ over the closed path between the extrema
$[r_{min},r_{max}]$, defined by the radii at which $\dot r=0$.  Thus,
\begin{equation} 
\beta_{w} \equiv  \oint_{r_{min}}^{r_{max}}\,d\tau =
\oint_{r_{min}}^{r_{max}}\, 
{dr \over \sqrt{ f \Big(f  - (2 \pi \sigma r)^2 \Big)}}.
\label{unchargedB} 
\end{equation} 
The reader is cautioned that this expression for the period
is inapplicable to the static, $m \to (6\sqrt{3} \pi \sigma)^{-1}$ case
in that as $r_{min} \to r_{max}$, we find $\beta_{w} \to 1/\sigma$.
This solution does not correspond to the static, $r= 3m$ solution 
for which the period in $r$ is infinite.
 
In order for solutions of the Euclidean equations of motion to define
{\it bona fide} domain walls, they should not intersect themselves.
Therefore the period $\beta_w$ must equal, or 
be an integer submultiple, of the
period of Schwarzschild:  
\begin{equation} 
\beta_{w} = {{\beta}_{S}\over n} = 
{8\pi m \over n}, \qquad n \in  {\Bbb Z}_{+}.  
\label{periodicity}
\end{equation} 
However, the above condition cannot be
satisfied for an accelerating black hole. There is no value of the
black hole mass in the interval $m \in [0, (6 \sqrt{3} \pi \sigma)^{-1}]$ 
for which the periods match, as shown in figure \ref{figure3}.  
It follows that the only non-self-intersecting
domain wall trajectory is the static one, in which case 
$r= 3 m$ and $m = 1/(6 \sqrt{3} \pi \sigma)$. Hence,
for a given domain wall surface energy density,
there is a unique mass for the Schwarzschild black hole which 
may be created.

We now calculate the action for the pair creation of static
black holes. From our earlier work, equation (\ref{actionderived}) in section
\ref{domains},
the Euclidean action is
\begin{equation} 
S_{E} = -{1 \over 2} \sigma \int_{W}\sqrt{h} d^3 x 
= -{1 \over 2} \sigma \,{\rm vol}(W),
\label{action1}
\end{equation} 
where $W$ denotes the hypersurface supporting the domain wall. 
Provided the surface energy density of the domain wall is positive,
the Euclidean action is {\sl negative}.
The volume of the domain wall $W$ is given by
\begin{eqnarray} 
{\rm vol}(W) &=&
\sqrt{1 - {2 m \over r}} \int_0^{8\pi m}  d\tau \int_0^\pi   d\theta 
\int_0^{2\pi} d\phi  \,  r^2   \sin\theta \cr\cr
&=& \sqrt{1 - {2 m\over r}} \cdot  8\pi m  \cdot 4\pi r^{2}  \cr\cr
&=& {96 \sqrt{3}} {\pi}^{2} m^{3}.
\end{eqnarray} 
Thus, the action
for the creation of a (static) Schwarzschild-domain wall
universe is given by 
\begin{equation}
S_{E} = -48 \sqrt{3} {\pi}^{2} m^{3}\sigma .
\label{wallaction}
\end{equation} 
Of course, we do not really want to create a whole new
universe containing a pair of static black holes and a
domain wall. Rather, we wish to calculate the probability
for black holes to form {\sl given} the presence of a domain
wall. We want to divide out by the probability
to nucleate a domain wall, as calculated in section \ref{domains}.  

The probability $P$ for the tunnelling process is the square of the
amplitude, $P= e^{-S_{E}}$.  Dividing the probability for creating
black holes with domain walls by the probability for domain walls only,
we obtain the relative probability for static, uncharged  black hole
pair production in the presence of a domain wall:
\begin{equation}
P = \exp\Big(-{11 \over 216 \pi\sigma^2}\Big).
\label{probuncharged}
\end{equation}
As expected, the probability is heavily suppressed for small $\sigma$.

This conclusion seems to be at variance with that of Hiscock
\cite{HISCOCK}, who has claimed that the Euclidean action for some
$O(3)$-symmetric bubble solutions is smaller than those of the
$O(4)$-symmetric solutions, and hence that black holes may act as
effective nucleation centers rendering the nucleation of domains more
likely. It seems that Hiscock is not considering precisely the same
situation that we have in mind, since we are considering a situation
in which the black holes appear at the same time as does the domain
rather than being present beforehand. Moreover, Hiscock considers the
general case when the energy densities $\rho_1,\,\rho_2$ outside the
domain wall are non-vanishing. It seems that one cannot obtain in a
simple way our solutions as a limit of his results. Thus, for
example, his equation (27) \cite{HISCOCK} is not equivalent to our
equation (\ref{wallaction}) in the limit that $\rho_1=\rho_2=0$.
This is for the same reason that (\ref{unchargedB}) cannot be used in
the limit $r_{min} \to r_{max}$ to obtain the static domain wall
solution.

Our results on the spatially-inhomogeneous domain wall space-time
seem to be in general accord with those of Bousso \& Hawking
\cite{BOUSSO} who consider spatially-homogeneous, complex solutions
with a massive scalar field. They find that the probability of
creating a Nariai-type solution with spatial topology $S^1 \times
S^2$ is suppressed relative to the probability of creating a de
Sitter-type solution with spatial topology $S^3$.

It is interesting to compare our result for the probability
(\ref{probuncharged}) with the rate for the pair creation of black
holes from a cosmic string.  In Eardley et al \cite{EARDLEY}, the
probability that a cosmic string will snap, producing two black holes
on the bare string terminals, was estimated. For a string of
mass-per-unit-length $\mu$ producing black holes of mass $m$, the rate
is approximately $\exp(-\pi m^2/\mu)$. To compare this result with
(\ref{probuncharged}), we look for black holes of mass $m= (6\sqrt{3}
\pi \sigma)^{-1}$. The ratio of the Euclidean actions for these
processes is $S_{E,wall}/S_{E,string} \sim 11 \mu/2$, so that for
Planck scale defects only do the two processes have comparable
probabilities.

%%%%%%%%%%%%%%%%%%%%%%%%%%%%%%%%%%%%%%%%%%%%%%%%%%%%%%%%%%%%%%%%%%%
%%%%%%%%%%%%%%%%%%%%%%%%%%%%%%%%%%%%%%%%%%%%%%%%%%%%%%%%%%%%%%%%%%%
%%%%%%%%%%%%%%%%%%%%%%%%%%%%%%%%%%%%%%%%%%%%%%%%%%%%%%%%%%%%%%%%%%%
\section{Creation of Magnetically Charged Black Holes}
\label{charged}

In this section, we consider the creation of magnetically
charged black holes. The case of electrically charged black holes
is entirely analogous with the subtlety that the electromagnetic
field in the electrically charged case must be pure imaginary
on the Riemannian section \cite{ROSS}.  
In the case of the charged black hole, the equation of motion
of the bubble wall is  
\begin{equation}
\sqrt{ \widetilde f - \dot r^2 } = 2 \pi \sigma r
\end{equation}
where $\widetilde f = 1 - 2m/r + q^2/r^2$, the
$g_{00}$ Euclidean Reissner-Nordstr{\"o}m metric coefficient in
$(\tau,r)$ coordinates, and 
$\cdot = \widetilde f^{-{1 \over 2}} {\partial \over \partial \tau}$.  
As with the uncharged case, a static solution exists for the motion
of the domain wall relative to the black hole, now located at a radius
\begin{equation}
r_{static} = {3 \over 2} m \Big[ 1 + 
\sqrt{1 - {8 \over 9} {q^2 \over m^2}} \Big].
\label{staticchargedradius}
\end{equation}
The mass of the created black holes, given by
\begin{equation}
m = {1 \over 6 \sqrt{6} \pi \sigma}
\Big[ 1 + 36 ( 2 \pi \sigma q)^2 + 
\Big( 1 - 12 (2 \pi \sigma q)^2 \Big)^{3/2} \Big]^{1/2},
\label{staticchargedmass}
\end{equation}
as a function of the black hole charge, $q$, 
and surface energy density, $\sigma$, of the
domain wall. Here we see that the mass runs between 
$(6\sqrt{3} \pi \sigma)^{-1} \le m \le (8\pi \sigma)^{-1}$
for $0 \le q \le m$.

The Euclidean action for the instanton now includes
an electromagnetic contribution. Using the Einstein-Maxwell
field equations, we obtain
\begin{eqnarray} 
S_{E} &=& -{1 \over 2} \sigma \int_{W} \sqrt{h} d^3 x 
+  \int_M \sqrt{g} d^4 x \,{F^2 \over 16 \pi}.
\label{EMaction}
\end{eqnarray}
The first term, due to the presence of the domain wall, gives
\begin{equation}
-{1 \over 2} \sigma \int_{W} \sqrt{h} d^3 x  =
-2 \pi \sigma r^2 \beta_{RN} {\widetilde f}^{1/2} |_{r_{static}}
\end{equation}
evaluated at $r_{static}$. Here, $\beta_{RN}$ is the instanton
period for the
Reissner-Nordstr{\"o}m black hole,
\begin{equation} 
{\beta}_{RN} = 2\pi
\Big[ {(m + \sqrt{m^2 - q^2})^2 \over \sqrt{m^2 - q^2} }\Big].
\end{equation} 
For the second term, the integration over $M$ covers
both sides of the domain wall space-time, to give
\begin{equation}
\int_M \sqrt{g} d^4 x \,{F^2 \over 16 \pi}
= q^2 \beta_{RN} \Big( {1 \over r_{+}} - {1 \over r_{static}} \Big),
\end{equation}
where $r_{+} = m + \sqrt{m^2 - q^2}$ is the outer black hole horizon radius.
We may divide the amplitude for this process
by the amplitude for domain creation
to obtain the probability for the pair creation of static, charged
black holes in the presence of a domain wall:
\begin{equation}
P = \exp\Big[ -{1 \over 8 \pi \sigma^2} 
+ 2 \pi \sigma r_{static}^2 \beta_{RN} \widetilde f^{1/2}
- q^2 \beta_{RN} \Big( {1 \over r_{+}} - {1 \over r_{static}} \Big)
\Big] 
\end{equation}
In the limit $q\to m$, the probability becomes 
$P=\exp[-3/(32\pi \sigma^2)]$.
Hence, we see that the probability for the pair
creation of static, charged black holes is only slightly
suppressed relative to the production of uncharged black holes.
 
For the case of a charged, accelerating black hole
we must match the domain wall instanton to the Lorentzian space-time.
The period of the instanton is now given by
\begin{equation} 
\beta_{w}
\equiv \oint_{r_{min}}^{r_{max}}\,d \tau
=  \oint_{r_{min}}^{r_{max}}\,
{dr \over \sqrt{ \Big(1 - {2 m \over  r} +{ q^2 \over r^2} \Big)
\Big(1 - {2 m \over r}  + {q^2 \over r^2} - (2 \pi \sigma r)^2 \Big)}}.
\label{chargedB}
\end{equation} 
Matching the Riemannian to Lorentzian sections, we require
that the period $\beta_{w}$ must equal, or be an integer
submultiple of, the period of Reissner-Nordstr{\"o}m:
\begin{equation} 
{\beta}_{w} = {\beta_{RN} \over n} = {2\pi  \over n} 
\Big[ {(m + \sqrt{m^2 - q^2})^2 \over \sqrt{m^2 - q^2} }\Big], 
\qquad n \in {\Bbb Z}_{+}.
\label{qperiodicity}
\end{equation} 
The behavior of $\beta_w$ and $\beta_{RN}$ as functions
of $q$ are shown in figure \ref{figure4}.
As $q \to m$, $\beta_{RN}$ diverges, whereas $\beta_{w}$ approaches
a finite value. Examining figure \ref{figure4}, we see that for a certain
values of the mass, there are values of the charge such that
$\beta_{RN} \ge \beta_{w}$. Alternatively, for certain values of
the charge, there are black hole masses such that 
$\beta_{RN} \ge \beta_{w}$.
Hence, we find the interesting result that for a given 
domain wall surface energy density $\sigma$, there exists a family
of instanton solutions describing the pair creation
of accelerating
black holes with mass $m$ and charge $q_n$ for $n\in {\Bbb Z}_{+}$.
These instantons are not $SO(2)$-invariant, merely $D_{n}$-invariant,
where $D_{n}$ is the dihedral group; the symmetries of a polygon with $n$
sides. If $n$ is odd, the maximum and minimum values of $r$ are diametrically
opposite each other in the $r-\tau$ factor. A radial line passing
through both lies on a reflection symmetric axis, a possible nucleation
surface for the instanton. If $n$ is even, there are two types of
reflection symmetric axes, one passing through two opposite radial maxima,
$r_{max}$, and one passing through two opposite radial
minima of $r_{min}$. Thus, there
appear to be two possible nucleation surfaces for the instanton in this
case. 

The two nucleation surfaces for $n$-even have interesting physical
consequences. The Lorentzian data for the black holes formed at
$r_{min}$ describe decelerating black holes. These black holes form,
then collapse in on the domain wall. The data for the black holes
formed at $r_{max}$ describe accelerating black holes. These black
holes form, then accelerate away from the domain wall.

Furthermore, of these instantons, the $n-$even solutions may describe
the creation of non-orientable black holes \cite{CHAMBLIN} if points on
the Riemannian section which differ by a shift in $\tau$ of half a
period are identified together with reflection in the domain wall.

The Euclidean action for the instanton describing the pair production
of charged, accelerating black holes with a domain wall is again given
by equation (\ref{EMaction}) with the following changes.  The first
term, due to the presence of the domain wall, is 
\begin{eqnarray}
-{1 \over 2} \sigma \int_{W} \sqrt{h} d^3 x  &=& 
-2 \pi \sigma \int_0^\beta d\tau r^2 {\widetilde f}^{1/2} \cr\cr
&=& -4 \pi \sigma \int_{r_{min}}^{r_{max}} dr \,
{r^2 \over \sqrt{\widetilde f - (2\pi \sigma r)^2}}
\end{eqnarray} 
The second term, due to the electromagnetic field, is
\begin{eqnarray} 
\int_M \sqrt{g} d^4 x \,{F^2 \over 16 \pi} &=& 
q^2 \int_0^\beta d\tau \int_{r_{+}}^{r} {dr \over r^2} \cr\cr
&=& q^2 {\beta \over r_{+}} - 2 q^2 \int_{r_{min}}^{r_{max}}
{dr \over r} {1 \over \sqrt{ \widetilde f (\widetilde f - (2\pi\sigma r)^2)}}
\end{eqnarray} 
Dividing the amplitude for this process by the amplitude
for domain creation, we obtain the probability for the pair creation of
accelerating, charged black holes in the presence of a domain wall.
The argument of the exponent of the probability, as a function of $q=q_n$,
for different values of $n$, are shown in figure \ref{figure5}.  
Here we see that the probability for the creation of accelerating,
charged black holes decreases only slightly for increasing
values of $n$.

It is interesting to compare our results with the recent work of Mann
\& Ross \cite{MANN} who consider Reissner-Nordstr{\"o}m - de Sitter
solutions. They find three classes of instantons for the pair creation
of charged black holes in a background de Sitter universe.  The first
class corresponds to black holes with zero surface gravity. These are
analogous to our extreme, static solutions. The second class occurs if
the surface gravities of the cosmological and event horizons are equal.
As first pointed out by Mellor \& Moss \cite{MOSS}, this arises if
$q=m$ in the usual notation. This class seems to be analogous to our
non-extreme, static solutions.  The third class is a generalized Nariai
space-time. This has no obvious analogue in our work.

Because of the $O(3)$-invariance assumed by Mann \& Ross,
their Lorentzian solutions must, by Birkhoff's theorem, be static
and the Riemannian solutions invariant under $SO(2)\times SO(3)$.
Thus, it is impossible to find analogues of our accelerating
domain wall solutions which are invariant only under $O(3)$.

%%%%%%%%%%%%%%%%%%%%%%%%%%%%%%%%%%%%%%%%%%%%%%%%%%%%%%%%%%%%%%%%%%%
%%%%%%%%%%%%%%%%%%%%%%%%%%%%%%%%%%%%%%%%%%%%%%%%%%%%%%%%%%%%%%%%%%%
%%%%%%%%%%%%%%%%%%%%%%%%%%%%%%%%%%%%%%%%%%%%%%%%%%%%%%%%%%%%%%%%%%%
\section{Lorentzian Saddle Points}
\label{lorentzian}

In this section we would like to point out that there {\it is} a
Lorentzian solution describing a pair of black holes accelerating in a
domain wall background. As would be expected, this is obtained by
taking two copies of Schwarzschild, `cutting' each copy along the
surface of a hyperboloid, the three-surface of constant acceleration
surrounding each hole, identifying these
hyperboloids and then identifying the holes to compactify everything
(see figure \ref{figure2}).  This solution, which we call the
accelerating Schwarzschild-domain wall solution, can be used to
construct a `Lorentzian path' corresponding to the `birth' of a pair of
accelerating black holes in a domain wall background, {\it even though
no Euclidean instanton describing the process exists}.  To see how to
do this, first notice that there exists a well-defined involutive isometry, 
which we shall denote as $R_{T}$, on
each side of the wall, which in Kruskal coordinates $(T,Z,{\theta},{\phi})$
can be defined simply by taking $T \to -T$. This involution is obviously
well defined on the entire Schwarzschild-domain wall solution.
We remind the reader that
the Kruskal coordinates $T$ and $Z$, which cover the maximal extension
of Schwarzschild, are given in terms of the coordinates $r$ and $t$ by
\begin{eqnarray}
T &=& \sinh \left(\frac{t}{4m}\right) e^{\frac{r}{4m}} \sqrt{r - 2m} \cr\cr
Z &=& \cosh \left(\frac{t}{4m}\right) e^{\frac{r}{4m}} \sqrt{r - 2m}
\end{eqnarray}
Unfortunately, $R_{T}$ has $T = 0$ as a fixed point set.  In order to get
a freely acting involution we therefore need to compose $R_{T}$ with
some other map which does not have any fixed points.  Such a map is 
given, in terms of the natural Kruskal coordinates on each side of the
domain wall, by simply constructing a map corresponding to
`parity inversion':
\begin{equation}
{P: \,(T, Z, {\theta}, {\phi}) \,\longrightarrow\, (T, Z, {\pi} - {\theta}, 
{\phi} + {\pi})}
\end{equation}
By forming the composition, $PR_{T}$, of these two maps we 
therefore obtain a freely acting involutive isometry which acts on the
the whole accelerating Schwarzschild-domain wall solution.
On the domain wall itself, the map $PR_{T}$
restricts to the usual antipodal map on de Sitter space
\cite{GIBBONS1}.  If we identify this solution under the action of this
involutive isometry, we obtain a non-time and non-space orientable Lorentzian
manifold with a {\it single} boundary component homeomorphic to ${\rm
S}^{1} \,\times\, {\Bbb R}{\Bbb P}^{2}$. This is therefore an example of a
solution where we can find a Lorentzian `history' describing some
process, but there exists no Euclidean instanton. More precisely, there 
exists a Lorentzian `saddle point', which interpolates from `nothing'
(the empty set) to `something' (the desired Schwarzschild-domain wall
solution).  One might object that this construction does not in fact
yield the desired late-time solution since obviously the geometry is
neither space nor time-orientable.  However, this objection is not really
relevant for two reasons.  First of all, using \cite{CHAMBLIN} we know
that each side of the domain wall in the identified solution admits
the pin structure corresponding to the superselection sector of fermions
actually used in particle physics, and so we can consistently study 
solutions of the Dirac equation in this background.  Second of all,
as we have discussed in detail in \cite{CHAMBLIN}, whenever there 
exists some energy source which can contribute to decay processes
such as black hole pair production (using instanton techniques),
that same energy source can contribute to the birth of non-orientable
black holes.  Indeed, as we discuss in \ref{nonorientable} 
of this paper, many of the
instantons which we have constructed above to describe the production
of magnetically charged black holes can be identified to yield
non-orientable intantons which mediate the decay of vacuum 
domain wall solutions into solutions containing a pair of non-orientable
black holes.  The purpose of this section is to point out that there 
are some decay processes which cannot be mediated even by 
non-orientable instantons, but which {\it can} be described using
non-orientable {\it Lorentzian} manifolds.

Of course, we would
expect this sort of thing to happen in a variety of other situations as
well. For example, any description of gravitational kink-antikink pair
production would require the use of Lorentzian histories since by
definition, the `nucleation surface' for a gravitational kink could not
have positive definite signature.

An interesting point to make here is that very often the solutions which
can be `created from nothing' using instanton methods can also be 
identified under the action of some freely acting isometry to yield a 
Lorentzian saddle point which also describes the late time `creation'
event.  This statement is formalized in the following
\vspace*{0.3cm}

{\noindent\bf Fact:} Let $M^{L} = (M, g_{L})$ be any Lorentzian manifold 
admitting a Riemannian section $M^{R}$, and suppose that there exists
some freely acting involutive isometry on $M^L$.  Then it is possible to
find a Lorentzian saddle point which describes the `creation from nothing'
of $M^L$.

In other words, given a decay process mediated by an instanton one can
often find a Lorentzian path which corresponds to the same process.  To
see why this is true, recall that since there exists an instanton $M^R$
for $M^L$ we can match $M^L$ to $M^R$ across a spacelike three-surface
${\Sigma}$ of vanishing extrinsic curvature.  We can therefore cut
$M^L$ along ${\Sigma}$ and form the `double' $2M^L$ such that
${\Sigma}$ is the $t = 0$ fixed point set of time reversal $T: t
~{\longrightarrow}~ -t$ on $2M^L$.  Given the existence of another,
freely acting involution `F', we can form the freely acting isometry FT
and identify $2M^L$ under the action of this involution.  The resulting
manifold will then be a non-time-orientable Lorentzian manifold with a
single boundary component, such that the surface $t = 0$ will be the
initial data `created' by the instanton.  Whether or not the identified
Lorentzian path is space-orientable or not will depend upon whether or
not the freely acting involution is space-orientation preserving or
reversing.  In the above example with the domain wall, we were forced
to take $P$ as our freely acting isometry, and $P$ is space-orientation
reversing.  In other scenarios, it may be possible to choose
space-orientation preserving maps.  Such subtle differences will depend
upon the detailed geometry of the solution in question.

%%%%%%%%%%%%%%%%%%%%%%%%%%%%%%%%%%%%%%%%%%%%%%%%%%%%%%%%%%%%%%%%%%%
%%%%%%%%%%%%%%%%%%%%%%%%%%%%%%%%%%%%%%%%%%%%%%%%%%%%%%%%%%%%%%%%%%%
%%%%%%%%%%%%%%%%%%%%%%%%%%%%%%%%%%%%%%%%%%%%%%%%%%%%%%%%%%%%%%%%%%%
\section{Production of Kaluza-Klein Monopoles and the
Dynamics of Compactification}

In this section we shall give examples of the creation of Kaluza-Klein
monopole - anti-monopole pairs. Our construction has something in
common with the creation of pairs of monopoles by magnetic fields in
Kaluza-Klein (KK) theory \cite{KALUZAKLEIN}.

To begin with, consider the creation of a 5-dimensional universe
by a positive cosmological constant. By the obvious analogy with
the 4-dimensional case the instanton is $S^5$ with its standard
round metric. We write this as
\begin{equation}
ds_5^2 = d \tau^2 + \cos^2\tau d \Omega_4^2
\label{s5metric}
\end{equation}
where $d\Omega_4^2$ is the round metric on $S^4$. The entire 5-sphere
is obtained by allowing $-{\pi \over 2} \le \tau \le {\pi \over 2}$,
but for the real tunnelling geometry we have $-{\pi \over 2} \le \tau
\le 0$.  The Lorentzian section has $\tau = i t$ with $t > 0$. This
gives a 5-dimensional expanding de Sitter universe whose spatial
cross-sections are the round metric on $S^4$.  To obtain the
3-dimensional space and 4-dimensional space-time we must find a Killing
vector field $\partial/\partial x^5$ whose trajectories are generically
circles and implement the KK reduction procedure.

Whatever Killing field we choose it is clear from (\ref{s5metric}) that
both the internal dimensions and the 3 spatial dimensions will expand
exponentially at the same rate in the Lorentzian portion of the
geometry. Thus this  example is not very satisfactory from a physical
point of view but it does serve to illustrate the nature of the initial
creation process. In a more elaborate model one might introduce some
mechanism which would cause the scale of the internal dimension to
settle down to some fixed, small value.

In choosing the $U(1)$ Killing field we recall 
that in the language of \cite{GIBBONS2} $S^4$ may be regarded as
containing a NUT and an anti-NUT. In KK theory these NUT's
correspond to monopoles. Explicitly we express the round
metric on $S^4$ as
\begin{equation}
d\Omega_4^2 = d\rho^2 + \sin^2 \rho \Big[
{1 \over 4}(d\theta^2 + \sin^2\theta d\phi^2) + {1\over 4}
(d x^5 + \cos\theta d\phi)^2 \Big]
\label{s4metric}
\end{equation}
where $0 \le \rho \le \pi$ and $\theta, \phi,x^5$ are Euler angles on
$SU(2) \cong S^3$. The metric in the square brackets above is the standard
round metric on $S^3$ with unit radius. Thus $0 \le \theta \le \pi$, $0
\le \phi \le 2 \pi$, $0\le x^5 \le 4 \pi$.  The vector field
$\partial/\partial x^5$ generates the Hopf fibration on the 3-sphere
$\rho=$constant. This acts freely on $S^3$ but it 
has fixed points at the north and south poles of
$S^4$, $\rho = 0$ and $\rho = \pi$. To reduce to $4+1$ dimensions 
we write
\begin{equation}
ds_5^2 = {\rm e}^{-{4 \sigma \over \sqrt{3}}}
\Big( dx^5 + 2 A_\mu d x^\mu\Big)^2 +
{\rm e}^{ {2 \sigma \over \sqrt{3}}}
g_{\alpha \beta} dx^\alpha dx^\beta
\label{5metric}
\end{equation}
where $\sigma = \sigma(x^\alpha)$ is the ``modulus field'' and 
$g_{\alpha \beta}$ is the space-time metric. From equations
(\ref{s5metric},\ref{s4metric},\ref{5metric}) we have
\begin{eqnarray}
{\rm e}^{-{4 \sigma \over \sqrt{3}}} &=& {1 \over 4} \sin^2\rho
\cos^2 \tau \cr\cr
g_{\alpha\beta} dx^\alpha dx^\beta &=&
{2 \over \sin\rho \cos\tau}
\Big[ d\tau^2 + \cos^2\tau \Big( d\rho^2 +
{1 \over 4} \sin^2\rho ( d\theta^2 + \sin^2\theta d\phi^2)
\Big)\Big] \cr\cr
A &=& {1 \over 2} \cos\theta d\phi
\label{match5metrics}
\end{eqnarray}
The last expression above shows clearly that magnetic 
monopoles are involved. The spatial metric is
\begin{equation}
ds_3^2 = {2 \cos\tau \over \sin\rho} \Big[ d\rho^2 +
{1 \over 4} \sin^2\rho ( d\theta^2 + \sin^2\theta d\phi^2)\Big].
\end{equation}
Near $\rho = 0$ we have, with $r = 2 \sqrt{\rho}$
\begin{equation}
ds_3^2 \sim 2 \cos\tau   \Big[ dr^2 +
{r^2 \over 16}  ( d\theta^2 + \sin^2\theta d\phi^2)\Big].
\end{equation}
This metric is singular near the monopole at $r=0$,
because the area of a small 2-sphere of radius $r$ is
$\pi r^2/4$. Thus the monopoles are rather like global
monopoles in that they have a solid angle deficit (of $15 \pi/4$).

To understand better the physics going on here, recall 
that substituting the ansatz (\ref{5metric}) into the
5-dimensional Einstein action we get a 4-dimensional
action of the form
\begin{equation}
\int \sqrt{-g} d^4 x \, \Big[ R - 2 (\partial \sigma)^2
- {\rm e}^{-{2 \sigma\sqrt{3}}} F_{\mu\nu}^2
- 2 \Lambda_5 {\rm e}^{{2 \sigma\over\sqrt{3}}} \Big]
\end{equation}
where $\Lambda_5$ is the 5-dimensional cosmological constant. 
Thus, one expects the modular field $\sigma$ will roll down the
potential $\Lambda_5 {\rm e}^{{2 \sigma\over\sqrt{3}}}$,
which is precisely what equation (\ref{match5metrics}) shows,
because at late, real time $t = -i \tau$ we have as $t \to \infty$,
$\sigma \sim - \sqrt{3} t/4$.

The extension of this idea to the case of domain walls is now
immediate.  We replace the two solid 4-balls $B^4$ by two solid 5-balls
$B^5$.  Joining them together gives an almost everywhere flat ($AEF$)
$S^5$.  The spatial sections are $AEFS^4$. These admit an obvious
$SO(4) \subset SO(5)$ invariant action whose orbits are 3-spheres. We pick out
the $U(1)$ subgroup corresponding to the Hopf fibration. Explicitly, the
flat metric on $B^5$ can be written as:
\begin{eqnarray}
ds^2 &=& d\rho^2 + \rho^2 (d \tau^2 + \cos^2\tau d\Omega_3^2) \cr\cr
&=& d\rho^2 + \rho^2 \Big[ d\tau^2 + \cos^2\tau
\Big( {1 \over 4} (dx^5 + \cos\theta d\phi)^2 +
{1 \over 4} \sin^2\theta d\phi^2 + {1\over 4} d\theta^2 \Big)\Big].
\label{b5metric}
\end{eqnarray}
The Riemannian half of the geometry is
given by $-{\pi \over 2} \le \tau \le 0$. The Lorentzian section is given
by $\tau = i t,\, t>0$. Note that $t=\tau=0$, the initial nucleation
surface, corresponds to the flat 4-ball with metric
\begin{equation}
d\rho^2 + \rho^2
\Big( {1 \over 4} (dx^5 + \cos\theta d\phi)^2 +
{1 \over 4} \sin^2\theta d\phi^2 + {1\over 4} d\theta^2 \Big).
\label{b4metric}
\end{equation}
As is well known, this is a special case of the self-dual multi-center 
metrics used to construct KK monopoles. In other words, it may
be given by the form
\begin{equation}
ds^2 = V^{-1} (dx^5 + \vec\omega d\vec x)^2 + 
V d\vec x^2
\label{multimonopolemetric}
\end{equation}
with $\vec\nabla \times \vec \omega = \nabla V \Rightarrow \nabla^2 V = 0$.
The asymptotically locally flat (ALF) $k-$monopole metric corresponds to
\begin{equation}
V = 1 + \sum_{i=1}^{k} {1 \over | \vec x - \vec x_i|}.
\label{multimonopoleV}
\end{equation}
The asymptotically locally Euclidean (ALE) metrics which tend to the flat
metric are ${\Bbb R}^4/C_{k}$, $C_{k}$ being the cyclic group of
order $k$, and have the same form of the metric as (\ref{multimonopolemetric})
with the replacement $V \to V-1$ in (\ref{multimonopoleV}).
The flat metric (\ref{b4metric}) corresponds to $k=1$.

The modular field $\sigma$, which has no potential in this case,
is now given by
\begin{equation}
{\rm e}^{-{4 \sigma \over \sqrt{3}}} = {\rho^2 \over 4} \cosh^2 t
\end{equation}
and so at late times it rolls: as 
$t \to \infty, \, \sigma \sim -\sqrt{3} t/2$.
The three dimensional metric obtained  by reduction is
\begin{equation}
\Big( {1 \over \cosh^2 t} d \rho^2  + {\rho^2 \over 4}
(d\theta^2 + \sin^2\theta d\phi^2)\Big)
\end{equation}
The area of a small sphere of proper radius $r$ is
${\pi \over 4} \cosh^2 t$. The solid angular deficit is thus
$\pi (16 - \cosh^2 t)/4$ which starts off at $15 \pi/4$
as in the previous case but decreases with time, becoming negative
for $t>\cosh^{-1}4$.

The generalization to higher-dimensional reductions on a circle is
straightforward and follows closely the work in \cite{KALUZAKLEIN}.
If the dimensionality of the space-time is odd, one takes $S^{2 n + 1}$
as the instanton. The spatial sections are now $S^{2 n}$:
\begin{equation}
ds^2 = d\tau^2 + \cos^2 \tau d\Omega_{2 n}^2.
\end{equation}
Now consider spherical polar coordinates on $S^{2 n}$:
\begin{equation}
d\Omega_{2 n}^2 = d\rho^2 + \sin^2 \rho d\Omega_{2 n-1}^2.
\label{s2Nmetric}
\end{equation}
The group $SO(2 n)$ acts on $S^{2 n}$, fixing the north and
south poles, $\rho=0,\pi$. Its orbits are the $S^{2 n-1}$'s
given by $\rho=$constant. Now, pick the Hopf
$U(1) \subset SO(2 n)$ acting freely on $S^{2 n-1}$ as the KK circle.
The reduced spatial manifold is now of the form
\begin{equation}
d\rho^2 + \sin^2 \rho (d \Omega_{2(n-1)}^{FS})^2
\end{equation}
where the $d\Omega$ is the Fubini-Study metric on ${\Bbb C}{\Bbb P}^{n-1}$.
There are two singularities at the north 
and south poles which correspond to a Bais-Batenberg
monopole - anti-monopole pair \cite{BAISBATENBERG}. These cannot exist as
isolated objects because the metric of a single Bais-Batenberg pole is
not asymptotically flat; they exhibit a kind of confinement.
However a monopole - anti-monopole pair is allowed.

In the dimensionality of the space-time is even and one takes
$S^{2 n+2}$ as the instanton, the discussion is similar but
now in polar coordinates analogous to (\ref{s2Nmetric}) the sphere $S^{2 n}$
does not admit, for topological reasons, a non-vanishing
vector field. As explained in \cite{KALUZAKLEIN} one could use a $U(1)$
subgroup of $SO(2 n+1)$ acting on $S^{2 n}$
whose fixed point set is a circle. As $\rho$ varies we would
get a 2-brane of fixed points. Alternatively we could
Hopf-fibre the entire spatial $S^{2 n+1}$. Thus, 
the $(2 n+2)-$dimensional metric would look like
\begin{equation}
ds^2 = d \tau^2 + \cos^2\tau \Big[ (d\Omega_{2 n}^{FS})^2 +
(dx^5 + A)^2 \Big]
\end{equation}
for the Hopf 1-form $A$. Thus at $\tau=0$ a closed inflating
universe is born whose spatial sections have 
the geometry of ${\Bbb C}{\Bbb P}^n$ with its
Fubini-Study metric.

This model is especially interesting if 
$n=2$ because ${\Bbb C}{\Bbb P}^2$ is not the
sole boundary of any compact 5-manifold and so it cannot be born from
nothing in a non-singular way in a purely 5-dimensional theory, whether
Lorentzian or Riemannian. Our example shows that it could be born in a
6-dimensional KK theory in a perfectly non-singular way. Thus
KK theory may allow the evasion of no-go theorems derived using 
co-bordism theory.

Kaluza-Klein theory on a single $U(1)$ is not an especially attractive
theory but it serves to point the way forward to more elaborate examples.
Thus one could exploit the fact that $S^7$ is an $S^3$ bundle of $S^4$.
Taking the round metric on $S^8$ as an instanton one has
\begin{equation}
ds^2 = d\tau^2 + \cos^2 \tau \Big[ {1 \over 4} d\Omega_4^2
+ (\sigma^{a} + A^a_\mu dx^\mu)^2 \Big],
\end{equation}
where $d\Omega_4^2$ is the round metric on
$S^4 \cong {\Bbb H}{\Bbb P}^1$, $\sigma^a$, with $a=1,2,3$,
are left-invariant 1-forms on $S^3$, and 
$A^a_\mu$ in the Yang-Mills instanton field.
Thus at $\tau = 0$ a universe with $S^4$ spatial cross-section is born
together with an $SU(2)$ Yang-Mills instanton.

%%%%%%%%%%%%%%%%%%%%%%%%%%%%%%%%%%%%%%%%%%%%%%%%%%%%%%%%%%%%%%%%%%%
%%%%%%%%%%%%%%%%%%%%%%%%%%%%%%%%%%%%%%%%%%%%%%%%%%%%%%%%%%%%%%%%%%%
%%%%%%%%%%%%%%%%%%%%%%%%%%%%%%%%%%%%%%%%%%%%%%%%%%%%%%%%%%%%%%%%%%%
\section{Non-Orientable Black Holes}
\label{nonorientable}

As was mentioned briefly in \ref{charged}, it is possible to define certain
discrete involutive isometries on the instantons which mediate the
production of charged black holes in the presence of a domain wall,
such that if one identifies the instantons under the action of these
isometries the resulting manifolds are non-orientable.  It is easy to
see that these involutions on the instantons extend to well defined
maps on the Lorentzian sections and therefore that these `identified
instantons' mediate the nucleation of non-orientable black holes in the
presence of a domain wall.  We will now describe this construction in
more detail.

First of all, recall that from eq. \ref{qperiodicity} and figure
\ref{figure4} we have the fundamental result that for a given domain
wall surface energy density ${\sigma}$, there exists a countable
infinite set of instantons, where each instanton in the set mediates
the production of a pair of accelerating black holes of mass $m$ and
charge $q_{n}$.  We shall denote each orientable instanton as $M^n$.
The value of n determines the behaviour of the domain wall on the
instanton. It is very useful to visualize the motion of the domain wall
on the instanton.  To this end, imagine viewing the $({\tau}, r)$
`cigar' section of the instanton, illustrated in figure \ref{figure2}
`head on'. Viewed in this way, we see that when n is odd, the
domain wall will sweep out an `odd-leafed clover' on the surface of the
instanton.  When n is even, the clover will have an even number of
leaves.  Because the leaves of a regular even-leafed clover are always
diametrically across from each other, we see that there exists an
obvious isometry on the instanton:  Namely, if ${\beta}_{RN}$ is the
period of the instanton then the map defined by
\begin{equation}
I: ({\tau}, r) ~{\longrightarrow}~ ({\tau}+{\beta_{RN} \over 2}, r)
\end{equation}
is a discrete isometry.  However, this map is not freely
acting.  If we attempt to identify the instanton under the action of I
the resulting manifold will have singularities.  We therefore need
to find a freely acting involution and compose it with I to obtain
a freely acting involutive isometry.  Such a freely acting map is 
given by simply taking `parity inversion', which was introduced above
in section \ref{lorentzian} and which is given as
\begin{equation}
{P: \,({\tau}, r, {\theta}, {\phi}) \,\longrightarrow\, 
({\tau}, r, {\pi} - {\theta}, {\phi} + {\pi})}
\end{equation}
We therefore obtain a freely acting isometric involution on each even n
instanton by forming the map $PI$.  We can  identify each even n
instanton under the action of $PI$.  The resulting smooth manifolds
will be non-orientable.  In particular, the nucleation surface will now
have the topology $S^{1} ~{\times}~ {\Bbb R}{\Bbb P}^{2}$.  We shall
denote these identified instantons as $M^{n}_{PI}$, where the notation
is meant to read `the n-th instanton identified under the action of
$PI$'.

Of course, $P$ by itself is a perfectly respectable freely acting
involutive isometry and so we are free to identify {\it all} instantons
under the action of $P$.  In fact, we can even identify the odd n
instantons under the action of $P$ since we are no longer identifying
the period.  We shall denote these instantons as $M^{n}_{P}$, where
again the notation means `the n-th instanton identified under the
action of $P$.

One might now ask, what is the difference between $M^{n}_{PI}$ and
$M^{n}_{P}$, for a given even value of n?  Well, for one thing
$M^{n}_{PI}$ has exactly half the volume (and hence half the action)
that $M^{n}_{P}$ does.  It follows that the decay process mediated by
$M^{n}_{PI}$ is four times {\it less} likely as the
corresponding process mediated by $M^{n}_{P}$.  
We therefore see
that non-orientable black holes will be produced with less frequency
than their orientable counterparts.  
Furthermore, since the
instanton periods are different the corresponding Lorentzian sections
will have distinct thermodynamical properties. It would be interesting
to study the properties of these black holes further.

%%%%%%%%%%%%%%%%%%%%%%%%%%%%%%%%%%%%%%%%%%%%%%%%%%%%%%%%%%%%%%%%%%%
%%%%%%%%%%%%%%%%%%%%%%%%%%%%%%%%%%%%%%%%%%%%%%%%%%%%%%%%%%%%%%%%%%%
%%%%%%%%%%%%%%%%%%%%%%%%%%%%%%%%%%%%%%%%%%%%%%%%%%%%%%%%%%%%%%%%%%%
\section{Conclusion}
\label{conclusion}

In this paper we have studied  the production of pairs of neutral and
charged black holes by domain walls, finding classical solutions and
calculating their classical actions. We have found that neutral black
holes whose creation is mediated by Euclidean instantons must be
produced mutually at rest with respect to one another, but for charged
black holes a new type of instanton is possible in which after
formation the two black holes accelerate away from one another.  These
new types of instantons are  not possible in Einstein-Maxwell theory
with a cosmological constant. Another unusual property is that there
exist a countably infinite sequence of pair created charged,
accelerating black holes with charge $q_{n}$ for a given mass $M$ and
domain wall surface density $\sigma$. Surprisingly, the probability
amplitude for the creation of these pairs asymptotes, rather than
decays, for $q_{n} \to M$.

This process of black hole formation in the presence of a domain wall
is similar in spirit to the snapping of a cosmic string to form a pair
of black holes. However, true analog of the cosmic string process would
seem to be the puncture of a domain wall with a black string at the
boundary of the rupture. As this necessarily involves n-branes rather
than simply black holes, we have not considered such a process in this
paper.

We have also found that creation of non-orientable black hole solutions
can be mediated by Euclidean instantons and that in addition if one is
prepared to consider entirely Lorentzian no-boundary type contributions
to the path integral then mutually accelerating pairs may be created
even in the neutral case.

Finally we have considered the production of Kaluza-Klein monopoles
both by a standard cosmological term and in the presence of a domain
wall.  We obtain in this way a better understanding of the putative
process of "compactification" which plays a central role, albeit in a
slightly different context, in almost all recent unification attempts.
The main point is that the compactification process is accompanied by
the production of pairs of topological defects -- Kaluza-Klein monopoles in
the simplest case. Thus we have obtained a unified picture in which
three hitherto largely separate themes in quantum gravity :
\begin{enumerate}
\item the birth of the universe from nothing,
\item the production of primordial topological defects, and 
\item the existence of extra dimensions
\end{enumerate}
are brought together. It seems plausible that the basic ideas in 
this paper will extend to more complicated and hopefully more 
realistic examples.

%%%%%%%%%%%%%%%%%%%%%%%%%%%%%%%%%%%%%%%%%%%%%%%%%%%%%%%% 
\acknowledgements 

The work of RRC was supported by PPARC through grant number GR/H71550.
The work of AC was supported by NSF Graduate 
Fellowship No. RCD-9255644 (in Cambridge) and NSF PHY94-07194
(in Santa Barbara).
 
%%%%%%%%%%%%%%%%%%%%%%%%%%%%%%%%%%%%%%%%%%%%%%%%%%%%%%%% 

\begin{figure}
\caption{
The cut-and-paste construction of the domain wall space-time is
presented. In the top row, two copies of Minkowski space-time, depicted
by their conformal diagrams, are joined along an asymptotically-null
surface in each region, describing the location of the domain wall.  
The time-symmetric surface $T=0$ is shown on the diagrams. In
the middle row, the procedure for the construction of the instanton for
domain wall creation is depicted. Two flat $4$-balls are glued
back-to-back, yielding a `lens' owing to the ridge of curvature running
along the hemisphere at the location of the domain wall.  At the
bottom, the Riemannian space has been joined to the Lorentzian
space-time on the $T=0$ hypersurface, depicting the creation from 
nothing of Minkowski domains partitioned by a domain wall.
}
\label{figure1}
\end{figure}

\begin{figure}
\caption{
The cut-and-paste construction of the domain wall and Schwarzschild
black hole space-time is presented. In the first row, two copies of
Schwarzschild space-time are joined
along the $r = 3 m$ timelike hypersurface at the location of
the domain wall. In the second row, the identification of the two
external regions of the Schwarzschild space-times is shown.  This
yields a space-time with topology ${\Bbb R} \times S^1 \times S^2$. In
the third row, the construction of the domain wall and Schwarzschild
instanton is shown.  Two manifolds with topology $S^2 \times {\Bbb
R}^2$ are cut at the radius $r=3m$, and glued back-to-back, yielding
a surface with topology $S^2 \times S^2$ with a
ridge of curvature at the domain wall. In the fourth row, the
Riemannian space has been joined to the Lorentzian space-time on the
$T=0$ surface, depicting the creation from nothing of a closed
space-time containing two domain walls and two domains containing a
black hole in each.
} 
\label{figure2}
\end{figure} 

\begin{figure}
\caption{
The instanton period $\beta_{w}$ for uncharged, accelerating
black holes (solid line) as a function of black hole mass.
We see that there is no value of the mass for which 
$\beta_{w}$ matches the Schwarzschild period $\beta_S = 8\pi M$
(dashed line). 
}
\label{figure3}
\end{figure}

\begin{figure}
\caption{
The instanton period for charged, accelerating black holes,
as a function of charge. The period $\beta_{w}$ is shown for 
$2 \pi \sigma M = {27}^{-1/2}, 1/7, 1/10$
(the long dashed, short dashed, and dotted lines respectively).
The Reissner-Nordstr{\"o}m period $\beta_{RN}$ is shown (solid line).
Accelerating, charged black holes may be produced for discrete
values of the charge $q_n$ such that $\beta_{w} = \beta_{RN}/n$.
}
\label{figure4}
\end{figure}

\begin{figure}
\caption{
The probability $P$ for the creation of accelerating,
charged black holes as a function of charge. The probability
is shown for $2 \pi \sigma M = {20}^{-1/2}, {27}^{-1/2}, 1/10$
(the solid triangle, open square, and solid square respectively)
for discrete values of the charge $q_n$ for $n=1,2,5,10,20,50,100$.
The probability for the creation of static, charged black holes
is also shown (solid line).
}
\label{figure5}
\end{figure}

\end{document}